\documentclass[preprint]{aastex}
\newcommand{\mincir}{\raise
-2.truept\hbox{\rlap{\hbox{$\sim$}}\raise5.truept\hbox{$<$}\ }}
\newcommand{\magcir}{\raise
-2.5truept\hbox{\rlap{\hbox{$\sim$}}\raise4.truept\hbox{$>$}\ }}
\newcommand{\minmag}{\raise
-2.truept\hbox{\rlap{\hbox{$<$}}\raise6.truept\hbox{$<$}\ }}
\newcommand{\be}{\begin{equation}}
\newcommand{\ee}{\end{equation}}
\newcommand{\ba}{\begin{eqnarray}}
\newcommand{\ea}{\end{eqnarray}}
\newcommand{\brr}{\begin{array}}
\newcommand{\err}{\end{array}}
\newcommand{\bc}{\begin{center}}
\newcommand{\ec}{\end{center}}

\title{Cosmological Evolution of Linear Bias}
\author{S. Basilakos$^{1}$, M. Plionis$^{2}$}
\affil{$^1$ Astrophysics Group, Imperial College London, Blackett Laboratory,
Prince Consort Road, London SW7 2BW, UK}
\affil{$^2$ Institute of Astronomy \& Astrophysics, National Observatory of
Athens, Lofos Nimfon, Thesio, 18110 Athens, Greece}

\begin{document}

\maketitle

\section*{Abstract}
Using linear perturbation theory and the Friedmann-Lemaitre solutions of the 
cosmological field equations, we derive analytically a second-order
differential equation for the evolution of the 
linear bias factor, $b(z)$, between the background matter and 
a mass-tracer fluctuation field.
We find $b(z)$ to be a strongly dependent 
function of redshift in all cosmological models. 
%??For the Einstein
%de-Sitter model we find that the known $(1+z)$ dependence of the bias
%evolution is only part of the complete solution which depends also on 
%$\propto (1+z)^{3/2}$??.
Comparing our analytical solution with the semi-analytic model of Mo \& 
White, which utilises the Press-Schechter formalism and the gravitationally
induced evolution of clustering, we find an extremely good agreement 
even at large redshifts, once we normalize to the same bias 
value at two different epochs, one of which is the present.
Furthermore, our analytic $b(z)$ function agrees well
with the outcome of N-body simulations.
\vspace{0.25cm}

\noindent
{\bf Keywords} Cosmology: theory - large-scale structure of 
universe 

\clearpage

\section{Introduction}
The concept of biasing between different classes of extragalactic objects 
and the background matter distribution was put forward by Kaiser (1984)
and Bardeen et al. (1986) 
in order to explain the higher amplitude of the 2-point correlation function 
of clusters of galaxies with respect to that of galaxies themselves.

In this framework biasing is assumed to be statistical in nature;
galaxies and clusters are identified as
high peaks of an underlying initially Gaussian random density field. 
Biasing of galaxies with respect to the dark matter distribution was also 
found to be an essential ingredient of CDM models of galaxy formation in 
order to reproduce the observed galaxy 
distribution (cf. Davies et al. 1985; Benson et al. 2000). 

The classical approach to study the redshift evolution of bias 
utilises the ratio of the correlation functions of objects and dark matter, 
which are assumed to be related via the square of a scale independent
bias factor. However, in this study we will use the definition by which
the extragalactic mass tracer (galaxies, halos, clusters) 
fluctuation field, $\delta_{tr}$, is related 
to that of the underlying mass, $\delta_{\rm m}$, by 
\be\label{eq:1}
\delta_{\rm tr} = b \delta_{\rm m} \;\;,
\ee
where $b$ is the linear bias factor. Note that the former definition results
from the latter but the opposite is not necessarily true.
The bias factor may have many dependencies; even assuming that it is scale 
independent, it necessarily depends on the 
type of the mass tracer as well as on the epoch $z$, 
since the fluctuations 
evolve with time as gravity draws together galaxies and mass. 
It is evident, therefore, that the bias
factor should also depend on the different cosmological models and dark 
matter content of the Universe (for a recent overview see Klypin 2000).

More realistic biasing schemes have been proposed in the literature.
Coles (1993) introduced the idea of biased galaxy formation in which galaxies
form with a probability given by an arbitrary function of the local 
mass density. Mann, Peacock \& Heavens (1998) investigated the 
properties of different bias models of galaxy distributions 
that results from local transformations of the 
present-day density field. The deterministic and linear nature of 
eq.(\ref{eq:1}) has been challenged (cf. Dekel \& Lahav 1999; 
Tegmark \& Bromley 1999) and indeed 
some non-linearity of the biasing relation is necessary to reconcile high 
biasing with deep voids. 
Despite the above, the linear biasing assumption is still a useful first order
approximation which, due to its simplicity, it is used in most studies
of large scale (linear) dynamics (cf. Strauss \& Willick 1995 and references therein; 
Branchini et al. 1999; Schmoldt et al. 1999; Plionis et al. 2000). 

Different studies have indeed shown that the bias factor is a monotonically 
decreasing function of redshift. 
An important advancement in the analytical 
treatment of the bias evolution was the work of
Mo \& White (1996) in which they used the Press-Schechter (1974) formalism and
found that in
an Einstein-de Sitter universe the linear bias factor evolves 
strongly with redshift.
Using a similar formalism, Matarrese et al. (1997) extended the Mo \& White
results to include the effects of different mass scales (see also Catelan et al 1998).

Steidel et al. (1998) confirmed that the 
Lyman-break galaxies are 
very strongly biased tracers of mass and they found that 
$b(z=3.4) \magcir 6, 4, 2$, for SCDM, $\Lambda$CDM $(\Omega=0.3)$ and
OCDM $(\Omega=0.2)$, respectively (see also Giavalisko et al 1998). 
A similar value for the $\Lambda$CDM model was obtained by Cen \& Ostriker (2000)
using high resolution Nbody/hydro simulations in which they treated DM, gas as well
as star formation.
The use of high resolution N-body simulations (cf. Klypin et al. 1996; 1999, 
Cole et al. 1997 and references therein) have shown that anti-biasing 
($b< 1$) should exist at scales $r \sim 3 - 8 h^{-1}$ Mpc, for the open and flat 
low-$\Omega$ models, in contrast with $\Omega=1$ models, where $b>1$.   
Colin et al (1999), using high-resolution N-body simulations of
SCDM, $\Lambda$CDM, OCDM and $\tau$CDM models,
which avoid the so called ``overmerging" problem, 
found that indeed biasing evolves rapidly with redshift, while 
Kauffmann et al. (1999) combining semi-analytic models of galaxy
formation and N-body simulations has also studied
the evolution of clustering in different cosmologies.

In this paper we will not indulge in such
aspects of the problem but rather, working within the paradigm of linear and 
scale-independent bias, we will derive the functional form of
its redshift evolution in the matter dominated epoch 
and in all cosmological models. The
Einstein de-Sitter case has been studied in the past (cf. Nusser \&
Davis 1994; Fry 1996; Bagla 1998) using the continuity
equation, which is a first order differential equation, to derive
a solution, $\propto (1+z)$, valid only for low $z$'s. Our approach is to use the
perturbation evolution equation which combines the continuity, the
Euler and the Poisson equations and which is a second order differential
equation. We should therefore expect to find a further component to the
known solution.

The paper is organised as follows: in section 2 we discuss the basic models 
for the linear bias evolution, in section 3 we derive the basic
differential equation describing the evolution of the linear bias
factor, while in section 4 we present its analytical
solution for the different cosmological models and a comparison
with previous models and N-body simulation results.
Finally, in section 5 we summarise our main results.

\section{Models for bias evolution}
Theoretical expectations regarding the cosmological evolution of bias 
have been investigated using analytical calculations, semi-analytical 
approximations and N-body simulations.
In this section we shortly describe some of these models in
order to compare them with our results.
  
\subsection{Test Particle or Galaxy Conserving Bias (M1):} 
This model, proposed by Nusser \& Davis (1994), Fry (1996), Tegmark \&
 Peebles (1998), predicts the evolution of bias,
independent of the mass and the origin of halos, assuming only that the
test particles fluctuation field is related proportionally
to that of the underlying mass. 
Thus, the bias factor as a function of redshift can be written:
\be
b(z)=1+(b_{\circ}-1)D(z)^{-1} \;\; 
\ee
where $b_{\circ}$ is the bias factor at the present time. 
Bagla (1998) found that for SCDM model and in the range $0\le z \le 1$
the above formula describes well the evolution of bias.

\subsection{Halo Models (M2):} 
Mo \& White (1996) using the Press-Schecter formalism, have developed a model for 
the evolution of the correlation bias, which depends on halo mass, 
and found, in an Einstein-de Sitter Universe, that:
\be
b(z)=1+\frac{1}{\delta_c}\left[ \left(\frac{D(z_*)}{D(z)} \right)^2 -1
\right]
\;\;,
\ee
with $z_*$ a reference 
redshift \footnote{Consider the distribution of halos of mass $M$, or larger, 
before typical halos of this mass have collapsed. Therefore, to
quantify this, the parameter $\nu(M,z)=\delta_{c}/\sigma(M,z)$, which related 
directly with the bias, $b=1+\frac{1}{\delta_{c}}(\nu^{2}-1)$, is
usually utilized
(where $\delta_{c}=1.69$ is the critical overdensity for spherical collapse at $z=0$, and
$\sigma(M,z)$ is the rms linear mass fluctuation on the scale $M$ of halos linearly extrapolated
to redshift $z$) . Thus, $z_*$ is fixed by requiring $\nu(M,z_{*})=1$ (cf. Bagla 1998).}, 
$\delta_c$ the critical overdensity for a spherical top-hat collapse model and $D(z)=(1+z)^{-1}$ is  
the linear growth rate of clustering. Parametrising this equation 
to the present epoch one gets:
\be\label{eq:Mo}
b(z)=0.41+(b(0) - 0.41)D(z)^{-2}
\ee
Similarly, Matarrese et al. (1997) parametrising the evolution of bias 
for halos
above a certain mass $M$, obtain a similar expression for an Einstein-de Sitter Universe:  
\be
b(z)=0.41+(b_{\rm eff}(0) - 0.41)D(z)^{-\beta}
\ee
with $b_{\rm eff}$ the bias of a sample of halos with a range of masses and
$\beta$ depending on the minimum mass scale that contributes 
to the halo correlation function (with $\beta < 2$).

\section{Basic Equations}
The central issue here is to derive the basic differential equation 
which describes the evolution of bias. The present  
analysis is based on linear perturbation theory 
in the matter dominated epoch (cf. Peebles 1993) and it is an extension
of the M1 model.

The time evolution equation for the mass density contrast, 
$\delta_{\rm m}=(\delta\rho/\rho)_{\rm m}$, modelled as a 
pressureless fluid with general 
solution of the growing mode: $\delta_{\rm m} \simeq A(x)D(t)$, is
(cf. Padmanabhan 1993):
\be\label{eq:6}
\frac{\partial^{2}\delta_{\rm m}}{\partial t^{2}}+2 H(t) 
\frac{\partial \delta_{\rm m}}{\partial t}=
4\pi G \rho_{\rm m} \delta_{\rm m}\;\; ,
\ee  

\noindent
Assuming for simplicity that the mass tracer population is conserved
in time, ie., that the effects of non-linear gravity and hydrodynamics (merging, 
feedback mechanisms etc) do not significantly alter the population mean,
then a similar evolution equation, containing in the right hand side
the gravitational contributions of all the perturbed matter, should 
be satisfied for $\delta_{tr}$ (see also Fry 1996; Catelan et al. 1998): 
\be\label{eq:7}
\frac{\partial^{2}\delta_{tr}}{\partial t^{2}}+2 H(t) 
\frac{\partial \delta_{tr}}{\partial t}=
4\pi G \rho_{\rm m} \delta_{\rm m} \;\; .
\ee  
Differentiating twice eq.(\ref{eq:1}) and using 
eq.(\ref{eq:7}) and eq.(\ref{eq:6}) we obtain:
\be\label{eq:10}
\ddot{b} \delta_{\rm m}+
2\left[\dot{\delta}_{\rm m}+H(t)\delta_{\rm m}\right] \dot{b}+
\left[\ddot{\delta}_{\rm m}+2H(t) \dot{\delta}_{\rm m} \right] b=
4 \pi G \rho_{\rm m} \delta_{\rm m}
\ee
Then from equation eq.(\ref{eq:10}), eq.(\ref{eq:6}) and $\delta_{\rm m} \simeq A(x) D(t)$
we have:
\be\label{eq:11}
\ddot{b}D(t)+2 \left[\dot{D(t)}+H(t)D(t) \right]\dot{b}+4\pi G \rho_{\rm m}
D(t)b=4\pi G \rho_{\rm m} D(t) \;\; .
\ee
In order to transform eq.(\ref{eq:11}) from time to redshift 
we use the following expression:
\be\label{eq:5}
\frac{dt}{dz}=-\frac{1}{H(z)(1+z)} \;\;,
\ee
where the Hubble parameter is given by:
\be\label{eq:3}
H(z)= H_{\circ} E(z)   
\ee
with
\be\label{eq:4}
E(z)=\left[ \Omega(1+z)^{3}+\Omega_{R}(1+z)^{2}+
\Omega_{\Lambda} \right]^{1/2}
\ee
and $\Omega= 8\pi G \rho_{o}/3H_{o}^{2}$ 
(density parameter),
$\Omega_{R}=(H_{o}\alpha_{o} R)^{-2}$ (curvature parameter),
$\Omega_{\Lambda}=\Lambda/3H_{o}^{2}$ 
(cosmological constant parameter) at the present time which satisfy
$\Omega+\Omega_{R}+\Omega_{\Lambda}=1$
and $H_{\circ}$ is the Hubble constant.

\noindent
Finally, the growing solution (cf. Peebles 1993) as a function of redshift is:
\be\label{eq:24}
D(z)=\frac{5\Omega E(z)}{2}\int^{\infty}_{z} \frac{(1+x)}{E^{3}(x)} 
{\rm d}x\;\;. 
\ee

\noindent
Therefore, as the time evolves with redshift, utilised eq.(\ref{eq:5}), 
eq.(\ref{eq:3}) eq.(\ref{eq:4}) and the relation
\be
4\pi G \rho_{\rm m}=4\pi G \rho_{\circ} (1+z)^{3}=
\frac{3H_{\circ}^{2}}{2} \Omega (1+z)^{3}\;,
\ee
then the basic differential equation for the evolution of the linear
bias parameter takes the following form:
\be\label{eq:gen2}
\frac{{\rm d}^{2} b}{{\rm d} z^{2}}-P(z)\frac{{\rm d} b}{{\rm d} z}+
Q(z)b=Q(z) \;\; 
\ee
with basic factors, 
\be\label{eq:ff1}
P(z)=\frac{1}{1+z} - \frac{1}{E(z)}\frac{{\rm d}E(z)}{{\rm d}z}
-\frac{2}{D(z)}\frac{{\rm d}D(z)}{{\rm d}z}
\ee
and
\be\label{eq:ff2}
Q(z)=\frac{3\Omega (1+z)}{2E^{2}(z)} \;\; .
\ee
It is obvious that the above generic form 
depends on the choice of the background cosmology. Thus,
the functional form which satisfies the general bias solution for 
all of the cosmological models is:    
\be\label{eq:31}
b(z)=y(z;\Omega;\Omega_{\Lambda})+1
\ee
where $y$ is the general 
solution of the homogeneous differential equation: 
\be\label{eq:gen222}
\frac{{\rm d}^{2} y}{{\rm d} z^{2}}-
P(z)\frac{{\rm d} y}{{\rm d} z}+
Q(z)y=0  \;\; .
\ee
Whereas it is obvious that the present theoretical approach takes into 
account the gravity field, it does not interact directly with the nature of 
the DM particles. 

\section{Bias Evolution in different Cosmological Models}
In this section using both eq.(\ref{eq:gen222}) 
and Friedmann-Lemaitre solutions of the cosmological field equations
we present the analytical solution of bias evolution 
for the Einstein-de Sitter, the open and low-density flat cosmological
models.

\subsection{Elements of the Differential Equation Theory}
Without wanting to appear too pedagogical, we remind the reader some basic elements of 
differential equation theory (cf. Bronson 1973). If one is able to find 
any solution $y_{1}$ of eq.(\ref{eq:gen222}), then a second linearly 
independent solution be found very easily.
Let the second solution can be written as  
\be\label{eq:gen322}
y_{2}(z)=y_{1}(z)u(z)
\ee
where $u(z)$ is to be determined. Inserting  eq.(\ref{eq:gen322}) into 
eq.(\ref{eq:gen222}) and remembering that $y_{1}$ satisfies the same 
equation, we find the following equation for $u(z)$:
\be
\frac{{\rm d}^{2} u}{{\rm d} z^{2}}+\left [\frac{ 2({\rm d}y_{1}/{\rm d}z)-
P(z)y_{1}(z)}{y_{1}} \right ] \frac{{\rm d} u}{{\rm d} z}=0  \;\; .
\ee
Integrating the above equation we have 
\be\label{eq:gen422}
\frac{{\rm d} u}{{\rm d} z}=\frac{\rm const}{y_{1}^{2}(z)} {\rm exp} \left ( 
\int_{z_{0}}^{z} P(x){\rm d}x \right)
\ee
where $z_{0}$ is an arbitrary initial point. A further integration of 
eq.(\ref{eq:gen422}) yields $u(z)$, and inserting this value into 
eq.(\ref{eq:gen322}), we obtain the second solution    
\be\label{eq:gen522}
y_{2}(z)={\rm const} \; y_{1}(z)\int_{z_{0}}^{z} \frac{{\rm d}x}{y_{1}^{2}(x)}
{\rm exp} \left ( 
\int_{z_{0}}^{x} P(t){\rm d}t \right) \;\; .
\ee
The Wronskian of the two solution $y_{1}$ and $y_{2}$ is 
\be
W(y_{1},y_{2})=y_{1}^{2} \frac{{\rm d}u}{{\rm d}z}=
{\rm const} \; {\rm exp} \left (\int_{z_{0}}^{z} P(x) {\rm d}x
\right)\;\; .
\ee
Thus the Wronskian never vanishes which implies that any general solution of 
eq.(\ref{eq:gen222}) is a linear combination ($y=c_{1}y_{1}+c_{2}y_{2}$) of the
fundamental set of solutions $y_{1}$ and $y_{2}$.

\subsection{Einstein - de Sitter Model}
In this case the basic cosmological equations are the following:
\be
E(z)=(1+z)^{3/2} \;,
\ee
while the growth factor of the linear density contrast is
\be\label{eq:15}
D(z)=\frac{1}{(1+z)}
\ee
and thus 
\be
P(z)=\frac{3}{2(1+z)}
\ee
and 
\be
Q(z)=\frac{3}{2(1+z)^{2}} \;\; .
\ee
It can be found that the function $y_{1}(z)=(1+z)=D^{-1}(z)$  
is a solution of the eq.(\ref{eq:gen222}) 
which is to be expected from the M1 model.
Therefore, we are looking for the second independent solution 
of the eq.(\ref{eq:gen222}). Thus
according to the procedure, described before we can calculate the second 
solution directly from eq.(\ref{eq:gen322}) 
$y_{2}(z)=(1+z)^{3/2}=D^{-3/2}(z)$. The general 
solution of the second order differential eq.(\ref{eq:gen222}) is 
the following:
\be\label{eq:sol}
y(z;1;0)= {\cal A} D^{-1}(z)+ {\cal B} D^{-3/2}(z) =
{\cal A} (1+z)+{\cal B} (1+z)^{3/2} \;\; 
\ee 
with general bias solution $b(z)=y(z;1;0)+1$. To this end this 
analysis generalise the M1 model in the sense that the added function 
$y_{2}(z)$ dominates the functional form of the bias evolution.
Of course in order to obtain partial solutions for $b(z)$ we need 
to estimate the 
values of the constants ${\cal A}$ and ${\cal B}$, which means that
we need to calibrate the $b(z)$ relation using two different epochs:
$b(z=0)=b_{\circ}$ and $b(z=z_{1})=b_{1}$.
Therefore, utilised both the above general bias solution   
and the latter parameters, we can give the expressions for the above 
constants as a function of $b_{\circ}$ and $b_{1}$:

\be\label{eq:alfa}
{\cal A}=\frac{ (b_{\circ}-1)D^{-3/2}(z_{1})-(b_{1}-1)}
{D^{-3/2}(z_{1})-D^{-1}(z_{1})} \;\; ,
\ee

\be\label{eq:beta}
{\cal B}=\frac{ (b_{1}-1)-(b_{\circ}-1)D^{-1}(z_{1})}
{D^{-3/2}(z_{1})-D^{-1}(z_{1})} \;\; .
\ee 
For ${\cal B}=0$ (M1 model) we obtain, as we should, ${\cal A}=b_{\circ}-1$. 

Our generalised solution does not suffer from limitations in the value
of $b$ (as does the M1 solution); $b$ can take values $>1$ and $<1$. It
is interesting to compare our generalised test-particle bias with
the more elaborate halo and merging models.
Since our approach gives a family of bias
curves, due to the fact that it has two unknown parameters, 
(the integration constants ${\cal A},{\cal B}$), 
we evaluate the latter by using Steidel et al. (1998) value of
the bias for Lyman break galaxies which gives for $\Omega=1$, 
$b(3.4) \simeq 7$. Inserting
this into Mo \& White (1996) model we obtain $b(0)=0.75$.
In figure 1 we compare our solution with the Mo \& White model 
and to our surprise we find an excellent agreement. 
This implies that the complete test particle bias solution is an
extremely good approximation to the more elaborate halo solutions which
takes into account, via the Press-Schechter formalism, the collapse of
different mass halos at the different epochs.

We further compare our analytic solution 
with N-body estimates provided by Colin et al. (1999) 
and Kauffmann et al. (1999). In figure 2 our model is represented by lines
while the numerical results with the different symbols. It is 
evident that our analytic function, normalized to two different epochs
of the numerical results, fits extremely well
the behaviour of the N-body derived bias evolution.

\subsection{Low Density Universes}
The basic cosmological equations in low-density
Universes become more complicated than in an $\Omega=1$ Universe
and eq.(\ref{eq:gen222}) does not have simple analytical solutions.  
We therefore present approximate analytical solutions 
which are valid in the high-redshift regime. In order to do so we
consider that {\bf (i)} for a low density open Universe, 
the Einstein-de Sitter growing mode is a good approximation 
for $z \magcir \Omega^{-1}-1$ 
and {\bf (ii)} for a low density flat 
Universe the growing mode is well approximated by the Einstein
de-Sitter case for $z \magcir \Omega^{-1/3}-1$
(cf. Peebles 1984b; Carrol, Press \& Turner 1992).

\subsubsection{Analytical Approximation for the Open Universe}
Using eqs (\ref{eq:ff1}), (\ref{eq:ff2}) and $z \magcir \Omega^{-1}-1$ 
we obtain the following 
basic factors of the differential equation (\ref{eq:gen222}):
\be\label{eq:ff12}
P(z)=\frac{2}{1+z} - \frac{\Omega}{2(1+\Omega z)}
\ee
and
\be\label{eq:ff22}
Q(z)=\frac{3\Omega}{2(1+z)(1+\Omega z)} 
\ee
where we have used $E(z)=(1+z)(1+\Omega z)^{1/2}$ and
eq.(\ref{eq:15}). In this case it can
be easily found that the function $y_{1}(z)=(1+z)+4(1-\Omega)/3\Omega$ 
is a solution of the eq.(\ref{eq:gen222}). 
Thus, after some calculations we can obtain the general 
bias solution: 
\be
b(z)-1=y(z \magcir (\Omega^{-1}-1);\Omega;0)
={\cal A}\left[(1+z)+4\frac{(1-\Omega)}{3\Omega}\right]+
{\cal C} \left[(1+z)+4\frac{(1-\Omega)}{3\Omega}\right] u(z)
\ee
with 
\be
u(z)=\int \frac{(1+z)^{2}\; {\rm d}z}{\left[(1+z)+4\frac{(1-\Omega)}{3\Omega}\right]^{2} 
(1+\Omega z)^{1/2}} \;\; .
\ee
Performing the latter integration one finds that the bias 
evolution is given by:
\be
b(z)-1 = {\cal A} \left[ (1+z)+4\frac{(1-\Omega)}{3\Omega} \right] 
+ \frac{2 {\cal C}}{\Omega} \left[ (1+z)+4\frac{(1-\Omega)}{3\Omega} \right] 
\left[ (1+\Omega z)^{1/2}+
\frac{8(1-\Omega)(1+\Omega z)^{1/2}}{(1-\Omega)+3(1+\Omega z)}\right]
\label {eq:sn}
\ee
It is obvious that for $\Omega \longrightarrow 1$ the above
solution tends to the Einstein-de Sitter case, as it should.

\subsubsection{Analytical Approximations for the $\Lambda$ Universe}

In a flat universe with non zero cosmological constant the growing
mode approximation leads to the following basic factors:
\be
P(z)=\frac{3}{1+z}-\frac{3\Omega (1+z)^{2}}{2[\Omega (1+z)^{3}+\Omega_{\Lambda}]}
\ee
and
\be 
Q(z)=\frac{3\Omega (1+z)}{2[\Omega (1+z)^{3}+\Omega_{\Lambda}]} \;\; 
\ee
where we have used $E(z)=[\Omega(1+z)^{3}+\Omega_{\Lambda}]^{1/2}$ and 
eq.(\ref{eq:15}). It is obvious that $y_{1}(z)=E(z)$ is a solution of
eq.(\ref{eq:gen222}) and following a similar procedure to that of the 
previous subsection we obtain the general solution:
\be
b(z)-1 = y(z \magcir (\Omega^{-1/3}-1); \Omega; \Omega_{\Lambda}) = 
{\cal A} \left[\Omega (1+z)^{3}+\Omega_{\Lambda}\right]^{1/2}+{\cal C} 
\left[\Omega (1+z)^{3}+\Omega_{\Lambda}\right]^{1/2} u(z)
\ee
where
\be\label{eq:88} 
u(z)=\int \frac{(1+z)^{3} {\rm d}z}
{[\Omega (1+z)^{3}+\Omega_{\Lambda}]^{3/2}} \;\; .
\ee
The integral of equation (\ref{eq:88}) is elliptic and therefore
its solution, in the redshift range $[z,+\infty)$, can be expressed 
as a hyper-geometric function. We finally obtain:
\be
b(z)-1={\cal A} [\Omega (1+z)^{3}+\Omega_{\Lambda}]^{1/2}+\frac{2{\cal C}}{\Omega^{3/2}} 
[\Omega (1+z)^{3}+ \Omega_{\Lambda}]^{1/2} (1+z)^{-1/2} 
F\left[\frac{1}{6},\frac{3}{2},\frac{7}{6},-\frac{\Omega_{\Lambda}}
{\Omega (1+z)^{3}} \right] \;\; . 
\ee
If $\Omega \longrightarrow 1$ and $\Omega_{\Lambda}\longrightarrow 0$ 
the above bias solution tends to the Einstein-de Sitter case, as it should.
Note that for $\Omega_{\Lambda}=0.7$ our solution is valid even at low redshifts, 
since the growing mode of the fluctuations evolution is well approximated by the 
Einstein de Sitter solution. Therefore we present in figure 3 a comparison 
between the results of the high-resolution N-body simulations 
of Colin et al. (1999) and our solution, parametrised to
$b(0)=0.75$ and the bias value of Colin et al. (1999) at $z=3$.
As it is evident the agreement is excellent.
 
\subsubsection{Analytic Solution for two Limiting Cases}
Although, due to the complex form of the $P(z)$ and $Q(z)$ functions
in the case of $\Omega<1$ Universes, we cannot solve the bias
evolution problem analytically for all redshifts,  
we can produce, however, a complete analytical bias evolution solution for the
limiting case of $\Omega=0$ (Milne Universe) or $\Omega=0$ and $\Omega_{\Lambda}=1$ 
(de Sitter Universe). 

In the former case we have from eqs (\ref{eq:ff1}) and (\ref{eq:ff2}) 
that $Q(z)=0$ and $P(z)=0$. Applying these to 
eq.(\ref{eq:gen222}), we find the general bias solution for
this special open cosmological model to be:
\be\label{eq:idiabias}
b(z)={\cal A}(1+z)+{\cal B}  
\ee
Interestingly, also the density fluctuations, $\delta$, have such a $z$-dependence
in the $\Omega=0$ Universe.
 
In the de Sitter Universe, which is dominated only by vacuum energy, 
putting  $\Omega \simeq 0$ into eq.(\ref{eq:ff1}) and 
eq.(\ref{eq:ff2}) 
we obtain again $Q(z)=0$, while the $P(z)$ factor is
given by: 
\be
P(z)=\frac{1}{1+z} \;\; ,    
\ee
with solution
\be
b(z)={\cal A} \int (1+z){\rm d}z+{\cal B}  \;\; .                        
\ee
Performing the latter integration one finds:
\be\label{eq:special2}
b(z)={\cal A} (1+z)^{2}+{\cal B}   \;\;,
\label{eq:nam2}
\ee
with initial condition $b_{\circ}={\cal A} +{\cal B}$.

We know that the two special solutions (\ref{eq:idiabias}) and 
(\ref{eq:special2}) do not correspond to realistic Universes. 
Nevertheless, these solutions can
operate as limiting cases of the generic problem. For example, using 
the general solution of the bias (eq.\ref{eq:31}) and 
putting $\Omega_{\circ}=0$, or $\Omega_{\Lambda}=1$ and
$\Omega_{\circ}=0$, the factors $(1+z)$ and $(1+z)^2$ 
should survive in the general case, as well.
In Figure 4, we compare the bias evolution 
of these special low-density models with the Einstein-de Sitter one, 
normalising them to the same $b(0)$ and using the results of
Steidel et al. (1998). Of course, by no means do we imply that these
models predict the same value of $b(0)$, we only parameterise our solution in 
order to compare the behaviour of $b(z)$ in these
limiting cases.

\section{Summary}
We have introduced analytical arguments and approximations based on linear 
perturbation theory and a linear, scale-independent bias between a mass
tracer and its underlying matter fluctuation field in order 
to investigate the cosmological evolution of such a bias.
We derive a second order differential equation, the solution of which
provides the functional form of the of bias evolution in any Universe.
For the case of an Einstein-de Sitter Universe, we find an exact solution
which is a linear combination of the known solution $\propto (1+z)$
(cf. Bagla 1998 and references therein), derived from the 
continuity equation, and a second term $\propto (1+z)^{3/2}$ which 
dominates. This solution once parametrised at two
different epochs, compares extremely well 
with the more sophisticated
halo models (cf. Mo \& White 1996) and with N-body simulations.

For the two low-density cosmological models we find exact solutions,
albeit only in the high-redshift approximation (where the
growing mode of perturbations can be approximated by the
Einstein-de Sitter solution).
We also derive analytical solutions for two limiting
low-density Universes (ie., $\Omega=0$, $\Omega_{\Lambda}=0$ and 
$\Omega=0$, $\Omega_{\Lambda}=1$).

\acknowledgments

We thank Stefano Borgani and Peter Coles for their useful comments and suggestions. 
We also thank the anonymous referee for the critical comments and useful 
suggestions. M.Plionis acknowledges the hospitality of
the Astrophysics Group of the Imperial College, were this work was completed.

\clearpage
\begin{figure}
\plotone{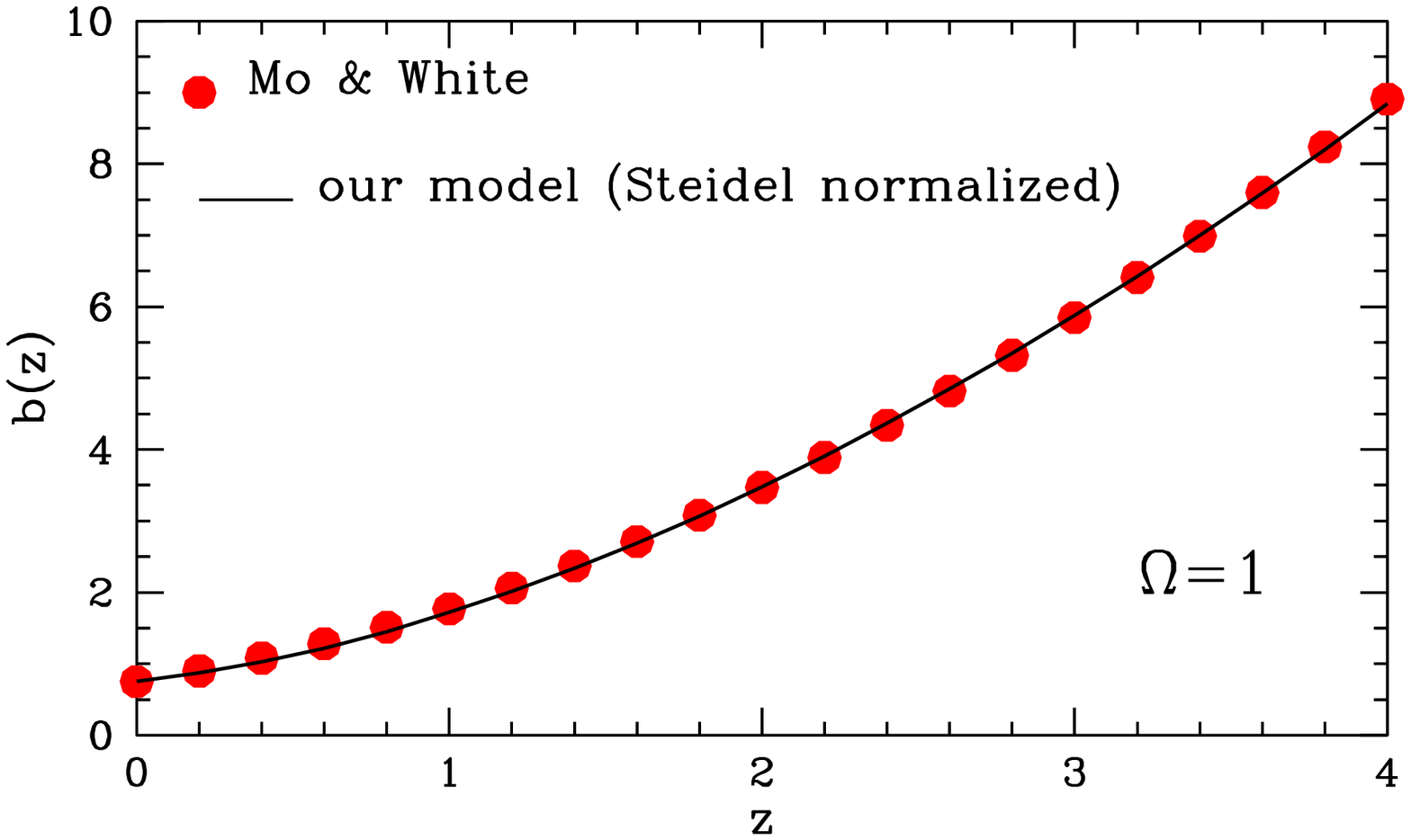}
\caption{Comparison of bias evolution between our solution (continuous line) and 
the functional form (eq.4) of the Mo \& White (1996) halo model (dots).}
\end{figure} 

\begin{figure}
\plotone{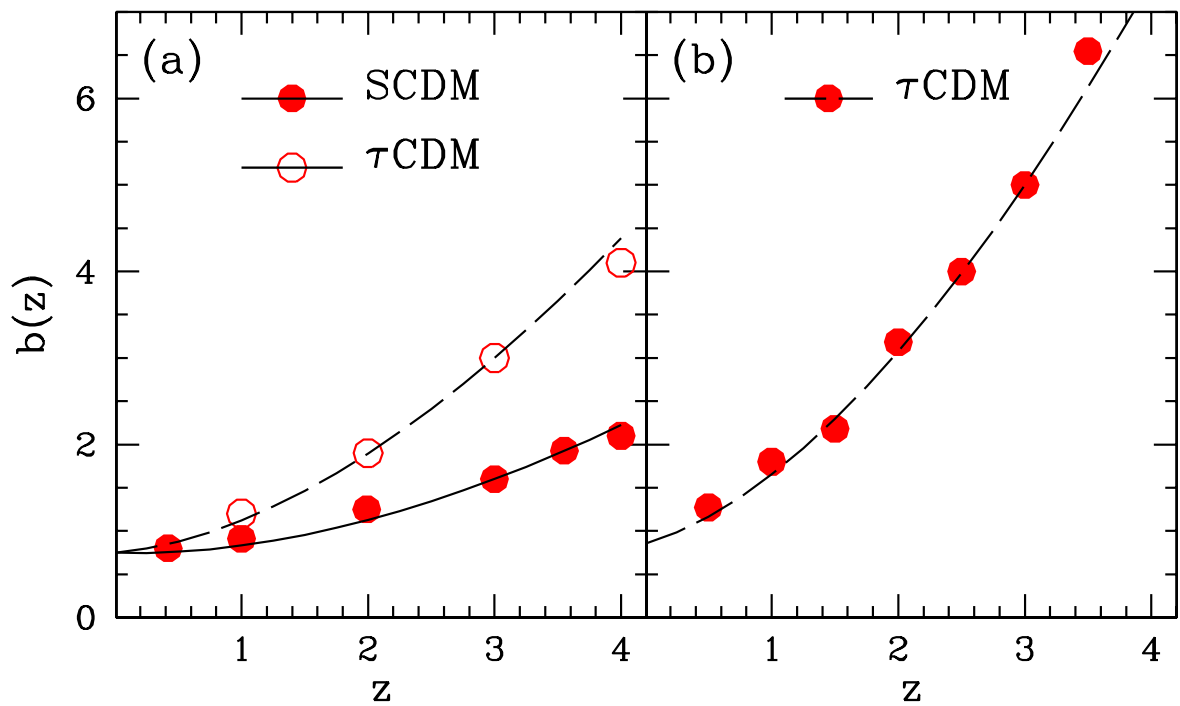}
\caption{Comparison of bias evolution between our solution and 
$\Omega=1$ N-body simulation results. (a) Lines represent
our model, parametrised to two epochs of the 
Colin et al. (1999) $\tau$CDM and SCDM models,
while points represent their results.
(b) Similarly for the Kauffmann et al. (1999) model.}
\end{figure} 

\begin{figure}
\plotone{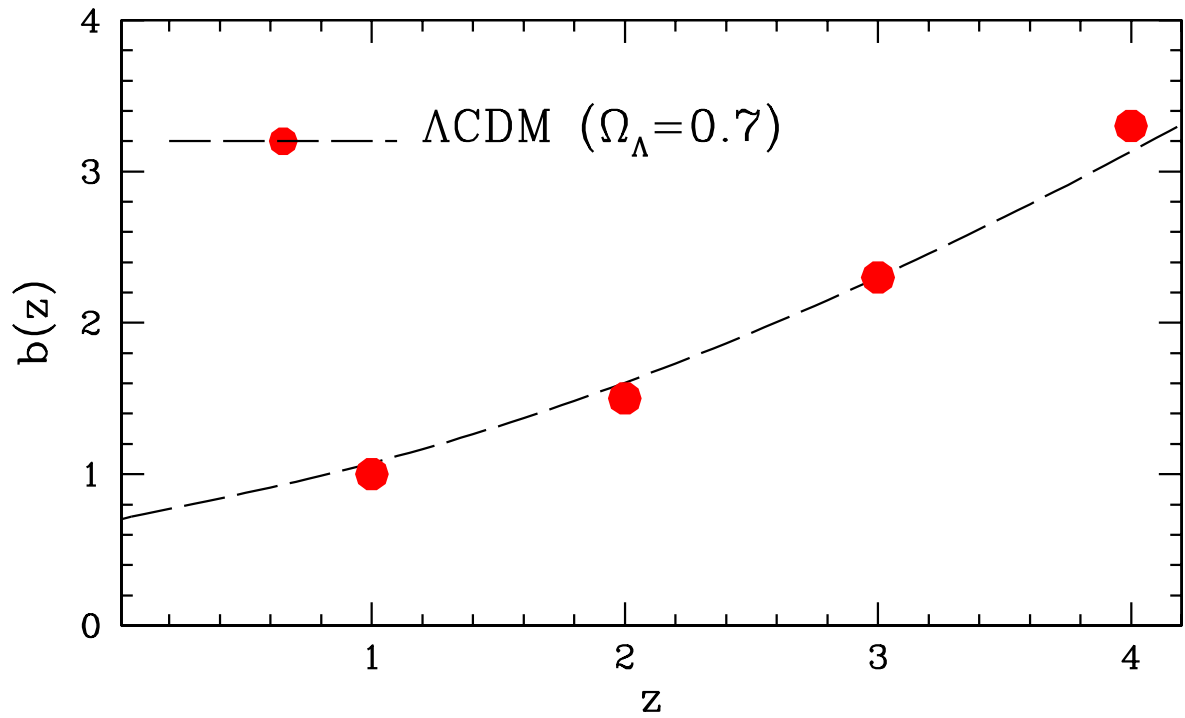}
\caption{Comparison of our $\Lambda$CDM $(\Omega_{\Lambda}=0.7)$ 
solution (broken line) 
with the results of high-resolution N-body simulations of
Colin et al. 1999 (points).}
\end{figure} 

\begin{figure}
\plotone{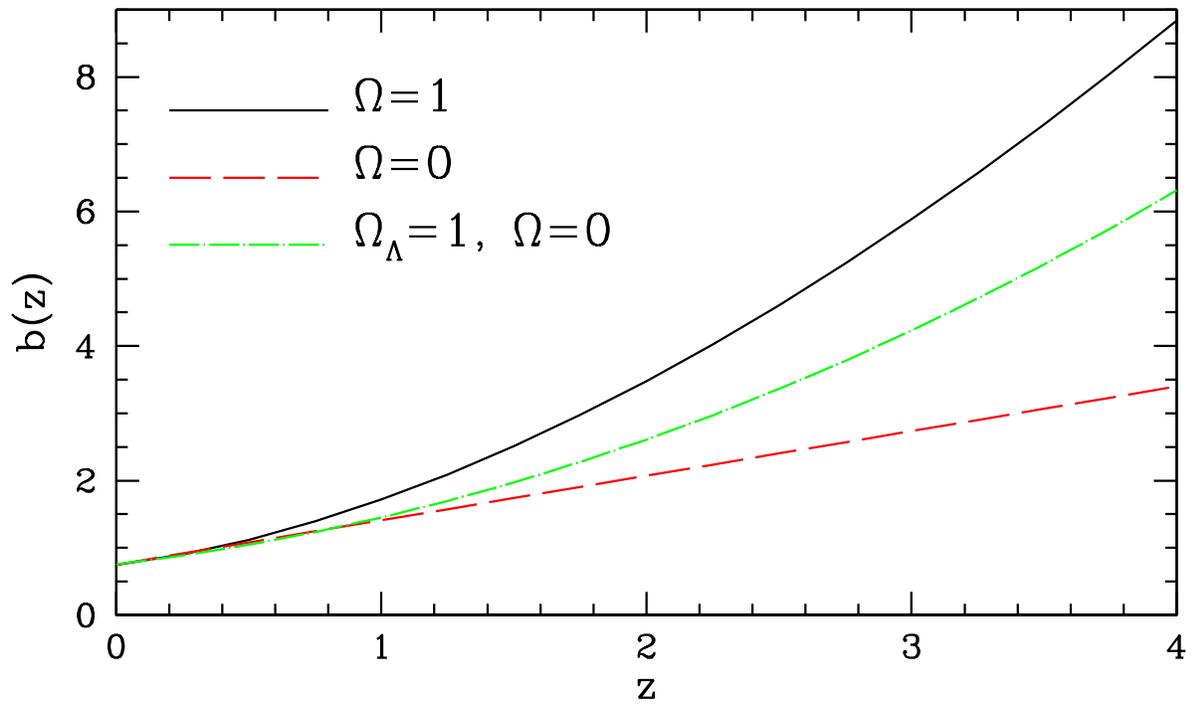}
\caption{Evolution of the bias factor for different models
parametrised using the  Steidel et al. (1998) bias values of
Lyman-break galaxies at $z=3.4$ ($b \simeq 6, 4, 2$, for SCDM, 
$\Lambda$CDM $(\Omega=0.3)$ and OCDM $(\Omega=0.2)$, respectively).}
\end{figure} 

\end{document}